\documentclass[twocolumn]{aa}
\usepackage{graphicx}
\usepackage{txfonts}
\usepackage{amsmath}
\usepackage{leftidx}
\usepackage{xcolor}

\usepackage{ulem}

\def\aap{Astron Astroph}
\newcommand{\dd}{\mathnormal{d}}
\newcommand{\old}[1]{}

\def\apjs{ApJS}

\def\aap{Astron Astroph}
\def\apjl{ApJ}

\newcommand{\betac}{\beta_{\text{c}}}

\DeclareUnicodeCharacter{2212}{-} 

\begin{document} 

\title{Magnetised tori in the background of a deformed compact object}

   \author{S. Faraji
          \and
          A. Trova 
          }

   \institute{University of Bremen, Centre of Applied Space Technology and Microgravity (ZARM), 28359 Germany\\
              \email{shokoufe.faraji@zarm.uni-bremen.de}\\ 
             \email{audrey.trova@zarm.uni-bremen.de}
                         }

   \date{Received ......; accepted .....}

\abstract
{This paper studied the relativistic accretion thick disc model raised by a deformed compact object that slightly deviated from spherical up to the quadrupole moment by utilising $\rm q$-metric. This metric is the simplest asymptotically flat solution of Einstein's equation with quadrupole moment. This work aims to study the effects of quadrupole moments in combination with the parameters of the thick magnetised disc model via studying the properties of these equilibrium sequences of magnetised, non-self-gravitating discs in this space-time. We employed different angular momentum distributions and discussed the procedure of building this toroidal disc model based on a combination of approaches previously considered in the literature. We have shown the properties of this relativistic accretion disc model and its dependence on the initial parameters. Besides, this theoretical model can be served as the initial data for numerical simulations.}


\maketitle

\section{\label{sec1}Introduction}

Studying accretion discs is currently a topic of widespread interest in astrophysics that links different areas of research. Given a general agreement, the observed properties of many astrophysical objects could be best explained in the framework of the accretion disc. However, there is no unified theoretical accretion disc model that could explain all the basic properties of these sources. Each of them only models some properties that are the best fit for the observation. The investigation of the proper disc models, by analytical or numerical setup, relies on constructing suitable representations based on physical assumptions. In this respect, analytic studies of equilibria and stability of fluid tori orbiting compact objects played a viable role in developing the black hole accretion theory.  


Among these successful theoretical models is the thick accretion disc with a toroidal shape, and no magnetic field was presented in these seminal works \citep{1980AcA....30....1J,1980A&A....88...23P} and followed by many authors. In particular \cite{1982Natur.295...17R} studied “ion tori” i.e. an optically thin, and ion pressure supported version of the thick disc model. Also \cite{1984MNRAS.208..721P} and
\cite{2006MNRAS.369.1235B} examined stability and oscillations of tori, also \cite{Komissarov_2006} constructed magnetised tori model.

On the other hand, there are studies where considering different distributions of angular momentum in the discs instead of setting this to be fixed. \cite{2009A&A...498..471Q} presented a method to build sequences of the black hole tori system in the dynamical equilibrium of the purely hydrodynamical case. Also, \cite{2015MNRAS.447.3593W} extended the original set of Komissarov' solutions in the presence of the particular case of power-law distributions of angular momentum, which is used in studying MRI instability through time-dependent numerical simulations \citep{2017MNRAS.467.1838F}. In this work\textbf{,} we generalised the extension of the models and combined the approaches described in \citep{Komissarov_2006}, \citep{2009A&A...498..471Q}, and \citep{2015MNRAS.447.3593W} to construct different magnetised disc models with non-constant angular momentum distributions, wherein this procedure, the location and morphology of the equipotential surfaces can be computed numerically.

Furthermore, most of the studies on accretion disc models considered accretion onto the Kerr black hole. However, in this paper, we study magnetised, perfect fluid tori in a particular static, axially symmetric metric that is non-spherically symmetric and characterised by two parameters: the mass M and the quadrupole moment q \citep{2011IJMPD..20.1779Q}. This metric is the simplest asymptotically flat solution of Einstein equation with quadrupole moment which describes the outer of a deformed compact object up to quadrupole and reads as

\begin{align}\label{Imetric}
	\dd s^2 &=- \left( \frac{x-1}{x+1} \right)^{(1+{\rm q})} \dd t^2+ M^2 (x^2-1)\left( \frac{x+1}{x-1} \right)^{(1+{\rm q})}\nonumber\\
	 &\left[\left( \frac{x^2-1}{x^2-y^2}\right)^{{\rm q}(2+{\rm q})} \left( \frac{\dd x^2}{x^2-1}+ \frac{\dd y^2}{1-y^2} \right)\right.\nonumber\\
	 &\left.+ (1-y^2) \dd{\phi}^2\right].\
\end{align}
Also, we compare the results of non-constant angular momentum with the constant angular momentum distributions in both cases of \textbf{a} spherical object and \textbf{a} deformed object.

There are some interesting reasons for studying the significance of the quadrupole moment for fluid equilibria and stability in this context, described e.g. by \cite{2005Ap&SS.300..127A}. Besides, there are other systems that can imitate a black hole's properties described in \citep{2002A&A...396L..31A,PhysRevD.78.024040,2019MNRAS.482...52S}. In general, this study can apply to the numerical simulations and explaining the observational data. Although in the more realistic scenario, the rotation should be taken into account; however, for example, it has been shown \citep{2004ApJ...617L..45B} that the possible resonant oscillations can be directly observed when arising in the inner parts of accretion flow around a compact object even if the source is steady and axisymmetric.


The organisation of the paper is as follows: the model is presented in \ref{sec3}. The results and discussion are presented in Section \ref{sec6}, and the conclusions are summarised in Section \ref{sec7}.
In this paper, the geometrised units where $c=1$ and $G=1$, also the signature $(-+++)$ are used.

\section{\label{sec3}The model}

The thick disc model presents a general method of constructing perfect fluid equilibria of matter in an axially symmetric and stationary space-time, which is the simplest analytical model of discs with no accretion flow, also radiatively inefficient. However, accretion rates can be very high ($\dot{m}\gg 1$), but the efficiency drops accordingly to very low values ($\eta_{\rm acc}\ll 1$). Indeed, this toroidal model is the relevant framework to describe properties of the astrophysical object, where the radial pressure gradients can not easily be negligible and their contribution leads to significant growth in the vertical size of the disc. 
Here we briefly mentioned the main assumptions and equations governing the magnetised model followed by \cite{Komissarov_2006}.

The evolution of an ideal magnetised fluid describes by the following conservation laws; baryon conservation, energy-momentum conservation, and induction equation \citep{1989rfmw.book.....A,1978srfm.book.....D}, respectively they read as

\begin{align}\label{eq:Property}
&\nabla_{\nu}\left(\rho u^{\nu}\right)=0\,,\\
&\nabla_{\nu}T^{\nu\mu}=0 \,, \\
&\nabla_{\nu}\leftidx{^*}{F}{^{\nu \mu}}=0\,,
\end{align}
where $T^{\mu \nu}$ is the total energy-momentum tensor of the fluid and electromagnetic field together, when the variations in pressure and density are adiabatic \citep{1989rfmw.book.....A},

\begin{align}
T^{\nu \mu}=\left(w+|b|^{2}\right) u^{\nu} u^{\mu}+\left(p_{\rm gas}+\frac{1}{2} |b|^{2}\right) g^{\nu \mu}-b^{\nu} b^{\mu},
\end{align}
here $w$ is the enthalpy, $p_{\rm gas}$ is the gas pressure, $u^{\mu}$ is the four-velocity of the fluid, and $b^{\mu}=(0, b^i)$ where $b^i$ are components of the three-vector magnetic field $B$. In fact, $b^{\mu}$ is related to the magnetic pressure in the fluid frame as $|b|^2=2p_{\rm m}$ \citep{1989rfmw.book.....A,1978srfm.book.....D}. Also, $\leftidx{^*}{F}{^{\nu \mu}}$ is the Hodge dual Faraday tensor

\begin{equation}
    \leftidx{^*}{F}{^{\nu \mu}}=b^{\nu} u^{\mu}-b^{\mu} u^{\nu}.
\end{equation}
Also, it is assumed to have purely rotational fluid motion and a purely toroidal magnetic field,

\begin{align}
u^x=u^y=b^x=b^y=0.
\end{align}
In fact, with these simplified assumptions, the only job is to solve the energy-momentum stress conservation \citep{1978A&A....63..221A,Komissarov_2006}. Adopting \citep{Komissarov_2006}, we assume the equations of state for fluid and for magnetic field as

\begin{align}
p=K w^{\kappa}, \quad \tilde{p}_{\mathrm{m}}=K_{\mathrm{m}} \tilde{w}^{\eta}
\label{eq:polytrops}
\end{align}
where $K$, $\kappa$, $K_{\mathrm{m}}$ and $\eta$ are constants. In fact, by considering these particular choices of polytropic equations of state, the von Zeipel theorem \footnote{The general relativistic version of the von Zeipel theorem states that for a toroidal magnetic field, the surfaces of constant $p$ coincide with the surfaces of constant $w$ if and only if constant $\Omega$ and constant $\ell$ coincide \citep{1924MNRAS..84..665V,2015GReGr..47...44Z}.}  is fulfilled, and by choosing specific angular momentum distribution, the final equation can be fully integrated.

\begin{align}\label{eq:FinalEq}
W-W_{\mathrm{in}}+\frac{\kappa}{\kappa-1}\frac{p}{w}+\frac{\eta}{\eta-1} \frac{p_m}{w}=\int_{\ell_{in}}^{\ell}{\frac{\Omega{\rm d}\ell}{1-\Omega\ell}},
\end{align}
where $W=\ln|u_t|$. So, by specifying $\Omega=\Omega(\ell)$, one can construct this model for $\Omega$ or $\ell$ profile and then $W(x,y)$ and $p(x,y)$ easily are followed. Also, one needs to specify $\ell(x,y)$ to fix the geometry of the equipotential surfaces.

\subsection{Angluar momentum profiles}
In this work we considered three different angular momentum distributions, namely the constant distribution as considered in \citep{Komissarov_2006}, the power-law distribution \citep{2015MNRAS.447.3593W}, and the trigonometric function \citep{2009A&A...498..471Q}.

\subsubsection{Constant distribution}
In the constant case, the right hand side of the equation \eqref{eq:FinalEq} vanishes and the disc surface is fully determined by the choice of $W_{\rm in}$ independent of magnetic field \citep{1989PASJ...41..133O}, and the value of $\ell_0$ determines the total potential

\begin{align}\label{K33}
    W(x,y)=\frac{1}{2}\ln|\frac{\mathcal{L}}{\mathcal{A}}|,
\end{align}
where $\mathcal{A}=g_{\phi\phi}+2\ell_0g_{t\phi}+\ell^2_0g_{tt}$. Then the gas pressure and magnetic pressure at the centre $c$, become

\begin{align}
    p_c= w_c(W_{\rm in}-W_{c})\left(\frac{\kappa}{\kappa-1}+\frac{\eta}{\betac(\eta-1)}\right)^{-1},
    \label{eq:pc}
\end{align}
where the subscript $c$ refers to the mentioned quantities at the centre. Also, the magnetisation parameter ${\betac}$ is the ratio of the gas pressure at the centre $p_c$ over the magnetic pressure at the centre $p_{mc}$. 

\subsubsection{Power-law distribution}

In the power-law distribution it is assumed that

\begin{align}
\Omega(\ell)=c\ell^{n}.    
\end{align}
In fact, this is often common to consider pure rotation and a barotropic equation of state. Thereafter, the equation \eqref{eq:FinalEq} can be written as \citep{2015MNRAS.447.3593W}

\begin{align}\label{eq:FinalEqw}
W-W_{\mathrm{in}}+\frac{\kappa}{\kappa-1}\frac{p}{w}+\frac{\eta}{\eta-1} \frac{p_m}{w}=\frac{1}{n+1}\ln\left(\frac{c\ell_{in}^{n+1}-1}{c\ell^{n+1}-1}\right).
\end{align}
In order to find $\ell$ one needs to calculate parameters $c$ and $n$. These are simply obtained by setting the places of $x_{\rm c}$, and $x_{\rm cusp}$.

\subsubsection{Trigonometric function distribution}

In this model it is assumed that \citep{2009A&A...498..471Q}

\begin{align}\label{triequ}
 \ell(x,y)=
   \left\{
  \begin{array}{@{}ll@{}}
  \ell_0\left(\frac{\ell_K(x)}{\ell_0}\right)^{\alpha}(1-y^2)^{\delta}, & x\geq x_{\rm ms}, \\
     \ell_0(\zeta)^{-\alpha}(1-y^2)^{\delta}, & x<x_{\rm ms},
    \end{array}\right.
\end{align}
where $\ell_0=\zeta \ell_K(x_{\rm ms})$, and $\ell_K$ is the Keplerian angular momentum in the equatorial plane. The parameters are determined by these bounds

\begin{align}
   0\leq \alpha\leq 1, \quad -1\leq \delta \leq 1, \quad -1\leq \zeta\leq \frac{\ell_K(x_{\rm mb})}{\ell_K(x_{\rm ms})}.
\end{align}
In this model one needs to solve 
\begin{align}
    \frac{\partial_xp}{\partial_{y}p}=\frac{\partial_xg^{tt}+\ell^2\partial_xg^{\phi\phi}}{\partial_{y}g^{tt}+\ell^2\partial_{y}g^{\phi\phi}}:= -F(x,y).
\end{align}
Therefore, the function $F$ is known once we have angular momentum distribution $\ell$.

In the next section, the results of considering these models combining with magnetized tori \citep{Komissarov_2006} are presented.



\section{\label{sec6}Results and discussions }
In this section, we discuss various models showing the impact of the deformation of the compact object due to quadrupole moments and the other parameters of the model on the location and morphology of the equipotential surfaces. 
We consider only slightly deformed compact objects in this work, namely small quadrupoles $\rm q$. Besides, the plots for various values of $\betac$ have been shown. However, in this work also the morphology of the solutions seems to have no change for the magnetisation parameter's values higher than $\betac=10^3$ or lower than $\betac=10^{-3}$. Also, these cases can thus be considered as the limiting cases in the simulations \citep{2017A&A...607A..68G}. Besides, the plots are presented in the $(x,y)$ coordinates.

\begin{figure}
    \centering
   \resizebox{\hsize}{!}{\includegraphics[width=7cm]{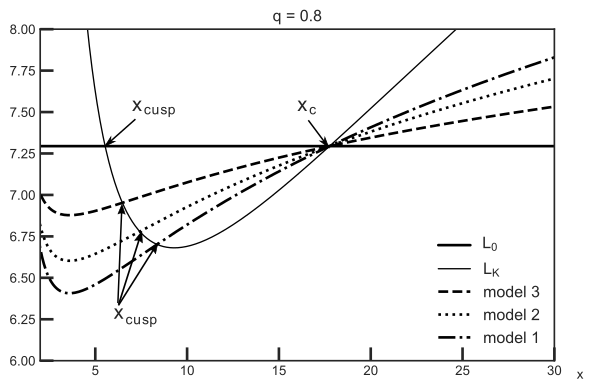}} 
    \caption{Various power-law angular momentum profiles are presented. The models $1$, $2$, and $3$ present different angular momentum profiles in the equatorial plane. The constant specific angular momentum is shown with the solid line which is chosen to be $\ell(x,y)=\ell_0=\ell_{\rm mb}(q)$. This profile corresponds to $\rm q=0.8$.}
    \label{fig:w1ModelL}
\end{figure}

In the case of power-law distribution, we have built three different angular momentum profiles via choosing three different values of $x'_{\rm cusp}$. The profiles are presented via numbers $1$, $2$ and $3$ in Figure \ref{fig:w1ModelL}, where through the paper, we have referred to these profiles simply only by model $1$, $2$ and $3$ for all the tested $\rm q$ values. In fact, the value of $x_{\rm cusp}$, $x'_{\rm cusp}$ and $x_{\rm c}$ depend on $\rm q$. As a result, the slope of the three models depends on $\rm q$ as well. Besides, the cusp and the centre are pushed away by increasing the value of $\rm q$. This may link to the fact that one can consider a positive quadrupole, models an oblate deformation on the central object, whereas a negative quadrupole models a prolate deformation.

\begin{figure*}[h]
\centering
\begin{tabular}{cc}
\includegraphics[width=7.5cm]{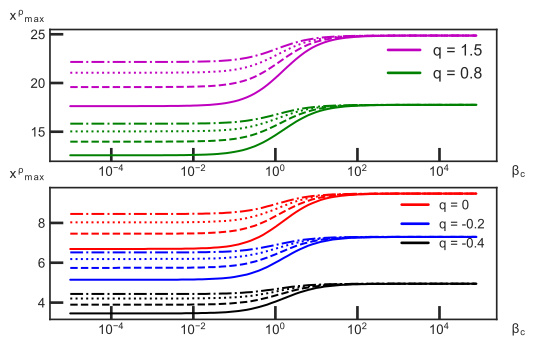}&
\includegraphics[width=7.5cm]{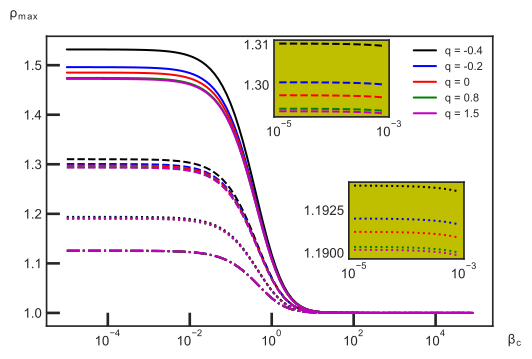} \\
\includegraphics[width=8cm]{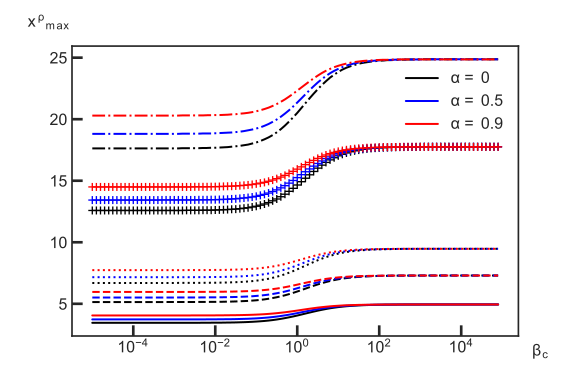}&
\includegraphics[width=8cm]{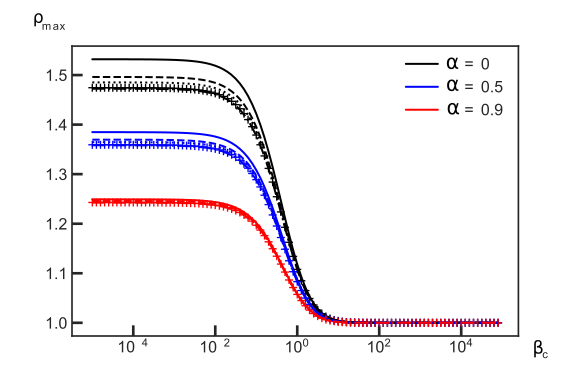} \\
\end{tabular}
\caption{Variation of the location (left panels) and amplitude (right panels) of the rest-mass density maximum over $\betac$ for different angular momentum profiles and $\rm q$. The top panels show the power-law distribution. The thick line shows the constant case, the dashed line presents model $3$, the dotted line model $2$, and the dot-dashed line model $1$ in Figure \ref{fig:w1ModelL}. The bottom panels present the trigonometric profiles of angular momentum in different colors. The dotted line presents $\rm q=0$, the solid line presents $\rm q=-0.4$, the dashed line $\rm q=-0.2$. Also, $\rm q=0.8$ and $\rm q=1.5$ are depicted in crossed line and dot-dashed line respectively.}
\label{fig:w2}
\end{figure*}

\begin{figure*}
    \centering
    \includegraphics[width=17cm]{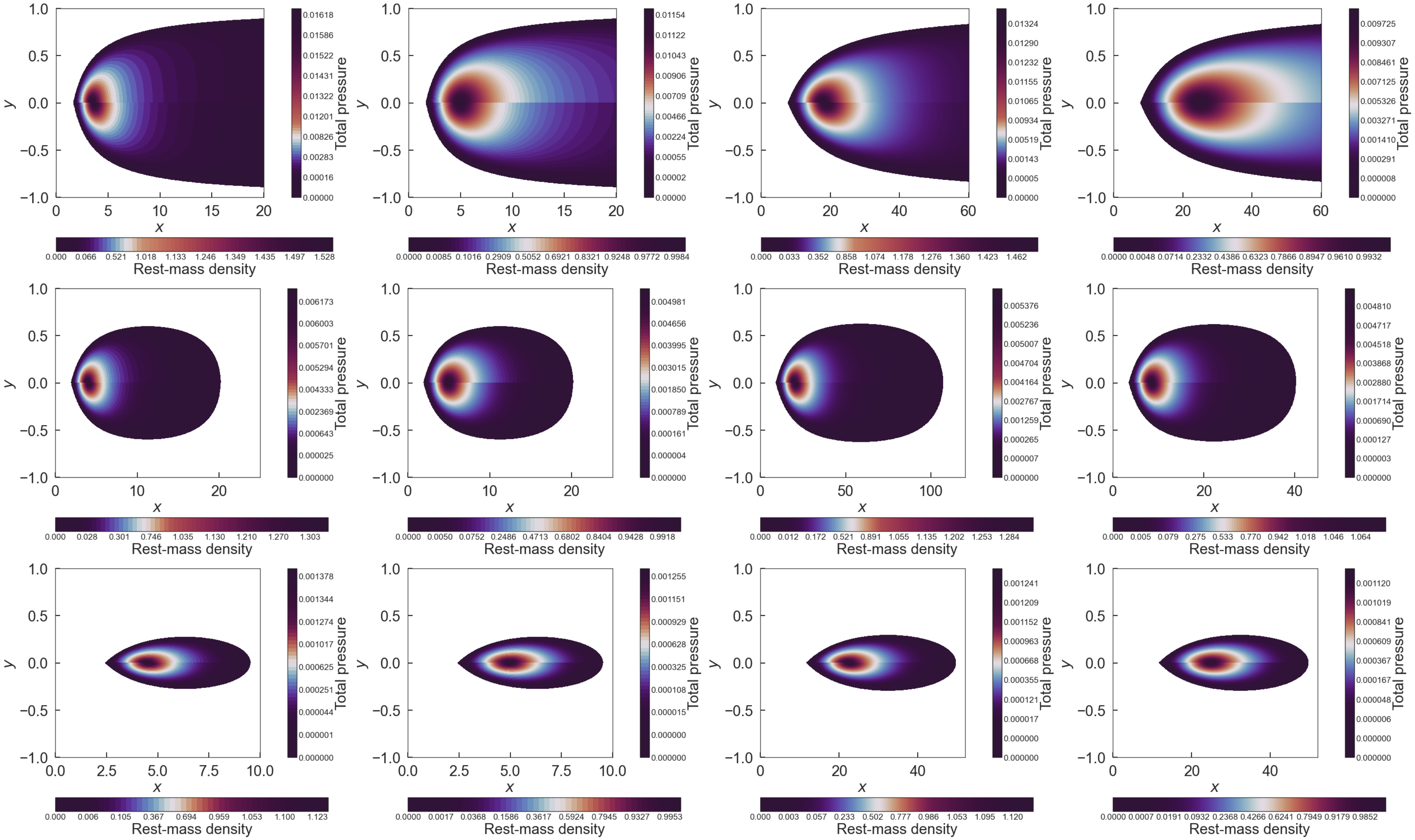}
    \caption{Map of the rest-mass density (top-half of each panel) and map of the total pressure (bottom half of each panel) for the power-law distribution. In the first two columns $q=-0.4$. In the last two columns $q=1.5$. Also, the column $1$ and $3$ present $\betac=10^{-3}$, and column $2$ and $4$ show $\betac=10^3$. Besides, each row depicts a different distribution of angular momentum, the constant case, model $3$, and model $1$, respectively.}
    \label{fig:MapW}
\end{figure*}

\begin{figure*}
    \centering
    \includegraphics[width=17cm]{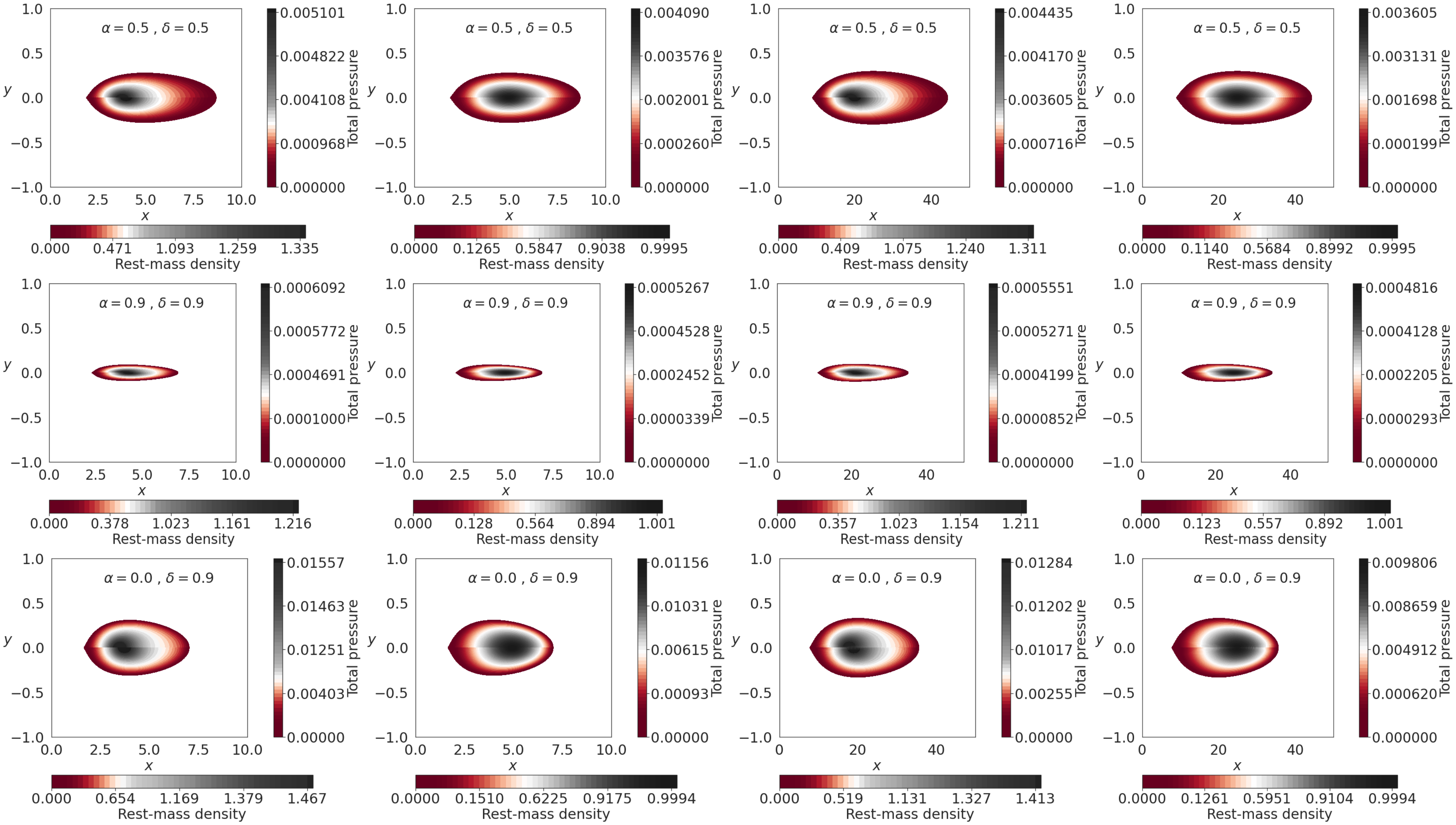}
    \caption{Map of the rest-mass density (top-half of each panel) and map of the total pressure (bottom-half of each panel). In the first two columns $q=-0.4$. In the last two columns $q=1.5$. Also, the column $1$ and $3$ present $\betac=10^{-3}$, and column $2$ and $4$ show $\betac=10^3$. The rows depict the effect of the different trigonometric angular momentum distributions.}
    \label{fig:MapQ}
\end{figure*}

In Figure \ref{fig:w2} top panels, the location and the amplitude of the rest-mass density maximum as a function of the magnetisation parameter for power-law distribution have been represented. In fact, for any fixed value of $\rm q$ and steepness of angular momentum profile, the location of the rest-mass density maximum by decreasing the value of $\betac$ -increasing the magnetic pressure- has been shifted inward. Also, if we concentrate on any chosen value of $\rm q$, we see the curves representing different models coincide when $\betac$ is very high. Indeed, when $\betac$ is low, the equidensity surfaces are not coinciding anymore with the equipotential surfaces. Also, decreasing $\betac$ increases the maximum of the rest-mass density. This causes to push the matter closer to the central object. Thus, the matter concentrates more on the inner part of the disc for any quadrupole value. This effect is coherent with the results of the previous studies \citep{Komissarov_2006,2017A&A...607A..68G}. This pattern also repeats for the trigonometric profiles of angular momentum distributions as seen at the bottom of Figure \ref{fig:w2}. In general, the magnetisation parameter has a major effect on how the matter is distributed in the disc compared to its effect on the geometrical structure of the disc that is more clearly seen in Figures \ref{fig:MapW} and \ref{fig:MapQ}.

Further, the location and amplitude of the maximum magnetic pressure follow the same path for maximum rest-mass density. The only difference is that the constant value reached by the pressure maximum's location depends on the angular momentum profile. An investigation on the different angular momentum profiles shows that the steeper the profile's slope, the farther the maximum of the rest-mass density from the central object. In fact, the higher steepness means that we have higher angular momentum at a given fixed place, as expected, leading to lower pressure. Then this is not a surprise that by using a steeper angular momentum profile, one obtains the maximum of the rest-mass density farther away from the central object; while having a decline in the maximum value of the density. This results in obtaining different shapes and sizes of the disc, which we see in the following Figure \ref{fig:MapW}. In fact, the steeper is the angular momentum profile, the less the disc is extended in both directions. The result is coherent with the fact that the steeper the profile, the closer $x_{\rm cusp}$ and $x_c$ to each other, as it has been seen clearly in the Figure \ref{fig:w1ModelL}. Further analysis reveals that as the steepness of the angular momentum profile increases, the place of pressure's maximum becomes closer to the centre of the disc rather than the cusp point. This is also expected since as the steepness increase, the distance between $x_{\rm cusp}$ and $x_c$ decreases; simultaneously, the place of density's maximum shifted to the outer part of the disc. So, this should become closer to the centre of the disc.

Thus, the angular momentum's steepness and magnetic pressure act contrariwise, or the angular momentum's steepness and magnetisation parameter $\beta_c$ act in the same way for all models and quadrupole's values. In fact, the steepness has a large effect on the density maximum and its place. Even when one chooses a profile with a high slope, it can neutralised the effect of $\rm q$ as seen in Figure \ref{fig:w2}. However, interestingly the effect's strength of $\beta_c$ depends on the slope of the angular momentum profile and the value of $\rm q$. This fact suggests a proper combination of the slope of angular momentum profile and value chosen for $\rm q$ can compensate the impact of magnetic pressure. In fact, this can be of great consequence, especially by applying to numerical simulation models.

\begin{figure*}
    \centering
    \includegraphics[width=6cm]{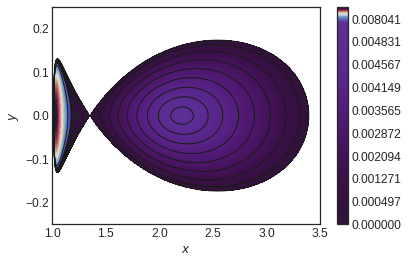}
     \includegraphics[width=6cm]{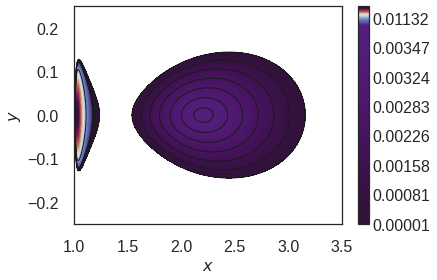}
    \caption{Possibility to have two tori in a row for $\rm q\in(-0.5,-0.553]$ for all discussed angular momentum distributions.}
    \label{fig:double}
\end{figure*}

Besides, an inspection on the impact of the parameter of geometry, deformation parameter $\rm q$, on the disc structure indicates that the negative values of $\rm q$ cause the place of the maximum of rest-mass density move inward and its value higher and vice versa for positive values.  We should mention in all considered angular momentum distributions, it is also possible to have connected or non-connected two tori in a row. This can happen when $\rm q\in(-0.5,-0.553]$. The equipotential surfaces in this case have been presented in Figure \ref{fig:double}. However, it needs further investigation to see if they present a physically meaningful model or not. In general, one can conclude that increasing $\rm q$ tends to make the disc shape thinner and more extended. It is reasonable to think that a prolate shape will enlarge the disc in the vertical direction closer to the inner edge. Also, an oblate form tends to flatten the disc on the equatorial plane. A comparison of all these parameters shows that the impact of $\rm q$ in the absence of the other parameters is less compared to the effect of slope of the angular momentum profile and $\beta_c$ on the disc's structure.

In the case of the trigonometric distribution, as shown in Figure \ref{fig:w2} at the bottom, we consider three different pairs of parameters. It is worth mentioning that the $\alpha$ parameter is the one more responsible for the location and amplitude of the rest-mass density and $\delta$ affects the distribution of equidensity surfaces; however, they can influence each other's impact, as we see in the following Figure \ref{fig:MapQ}. In this case, for small values of $\betac$, changing the profile of angular momentum has weaker effects on the displacement of maximum density and its position than changing the value of quadrupole. However, a deeper inspection shows that the strength of the impact of $\beta_c$ depends on both parameters $\rm q$ and $\alpha$. As for larger values of $\rm q$ and $\alpha$, the impact of changing in $\betac$ is tenser. Besides, for any fixed value of $\betac$ but not relatively large, the parameter $\alpha$ is also dependent on $\rm q$. They are all correlated with each other, especially when they are not very high or very low. Thus, by considering a proper combination of parameters, one can obtain the desired model, which could be applicable in fitting observational data. Of course, this work may consider as a first step and a foot-stone for simulation and numerical analysis; however, it requires further studies and considerations.




Further, we have built the solutions with this angular momentum ansatz for the rest-mass density and the total pressure in Figure \ref{fig:MapQ}. In this case,  $\betac$ does not influence the geometry of the disc as it was predicted. However, the matter would be more spread in the entire disc by increasing $\betac$. In fact, an analysis shows that the $\alpha$ parameter is more responsible for the vertical direction of the disc in the way that as much as we increase $\alpha$, we have a less vertically extended disc. While parameter $\delta$ affects both the radial and the vertical extension of the disc. Also, it affects the distribution of equidensity surfaces in general. However, the analysis on how $\delta$ affects the structure of the disc is in correlation with $\alpha$ that one can not study this contribution independently. 



\section{Summary and conclusion}\label{sec7}
This paper analysed equilibrium sequences of magnetised, non-self-gravitating discs around a deformed compact object up to the quadrupole. This solution is described via the $\rm q$-metric.

In this procedure we considered the three existing approaches of \citep{Komissarov_2006} combined with \citep{2009A&A...498..471Q}, also \citep{Komissarov_2006} combined with \citep{2015MNRAS.447.3593W} to study the space-time generated by this deformed object via analysing the properties of the magnetic tori disc model around this object. Also, the results are in good agreement with the result of the mentioned papers when we limited our attention to the vanishing quadrupole moment case. 

More precisely, in the case of the power-law distribution, we have shown that steeper angular momentum causes the disc to be more shrunk. Moreover, an increase in the steepness of the angular momentum profile decreases the amplitude of the rest-mass density and pushes away the location of its maximum. In fact, it tends to spread the matter in the entire disc rather than concentrate on the inner part.

In the case of the trigonometric distribution profile, we have shown that the pair ($\alpha$, $\delta$) affect strongly on the overall configuration of the disc. In particular, the $\alpha$ parameter is responsible to set the maximum rest-mass density and pressure, and their positions. Also, play a role in the vertical extension of the disc. While parameter $\delta$ affects the shape of the disc radially and vertically. However, in general, the effects of $\delta$ on the structure of the disc are in correlation with the parameter $\alpha$ in a way that one can not study this contribution independently.

Further, we have shown that changing parameter $\betac$ has a noticeable effect on the location and amplitude of the rest-mass density, also spreading the matter through the disc. In fact, a higher magnetic pressure, namely lower $\betac$, causes the matter to concentrate more in the inner part of the disc. Furthermore, in this case, the range of isodensity contours is increasing, which is compatible with the increase of rest-mass density in the inner part of the disc. Indeed, this results remain valid for any chosen value of quadrupole $\rm q$. In addition, the parameter $\rm q$ is more responsible for a shift of the whole disc, also its radial extension. Namely, the bigger we choose $\rm q$, the more extended disc we have. Also, for higher values of $\rm q$, the disc is more pushed outward from the central object. Besides, the rest-mass density maximum and its location are affected by quadrupole parameters; as such, the maximum is a decreasing function. Simultaneously, its location is an increasing function of $\rm q$. However, the effect of $\rm q$ is in correlation with the other parameters, in the sense that it always seems the strongest effect of different angular momentum profiles or different values of $\betac$ reveals for the smallest value of $\rm q$. As $\rm q$ increases, depending on the parameter values, one should consider their correlations to describe the disc model properly.
Moreover, there is a possibility of having two tori in a row for negative valid values for parameter $\rm q$, which needs further investigation to understand if they are physically reasonable results for astronomical systems.





As a further step of this work, the study on the oscillation of the disc in this setup is in progress. Also, one can investigate the other forms of the barotropic equation of state. It is also of some interest to apply these models as the initial conditions in the numerical simulations and test their ability to account for observable constraints of astrophysical systems.

\section* {Acknowledgements}
S.F. gratefully acknowledges Prof. Komissarov for his fruitful
comment. The authors are grateful to the anonymous referee for the useful comments. This work is supported by the research training group GRK 1620 ”Models of Gravity”, funded by the German Research Foundation (DFG).

\bibliographystyle{aa}
\bibliography{bibqtorishort}

\end{document}